# Roles of adiabatic and nonadiabatic spin transfer torques on magnetic domain wall motion

JAE-CHUL LEE[1,2†], KAB-JIN KIM[1†], JISU RYU[3], KYOUNG-WOONG MOON[1], SANG-JUN YUN[1], GI-HONG GIM[1], KANG-SOO LEE[1], KYUNG-HO SHIN[2], HYUN-WOO LEE[3‡], SUG-BONG CHOE[1*]

[1]Center for Subwavelength Optics and Department of Physics, Seoul National University, Seoul 151-742, Republic of Korea

[2]Center for Spintronics Research, Korea Institute of Science and Technology, Seoul 136-791, Republic of Korea

[3]PCTP and Department of Physics, Pohang University of Science and Technology, Pohang, Kyungbuk 790-784, Republic of Korea

[†]These authors contributed equally to this work.

[*]e-mail: sugbong@snu.ac.kr

[‡]e-mail: hwl@postech.ac.kr

**Electric current exerts torques—so-called spin transfer torques[1-2] (STTs)—on magnetic domain walls (DWs), resulting in DW motion[3-8]. At low current densities, the STTs should compete against disorders in ferromagnetic nanowires[9-15] but the nature of the competition remains poorly understood. By achieving two-dimensional contour maps of DW speed with respect to current density and magnetic field, here we visualize unambiguously distinct roles of the two STTs—adiabatic[1] and nonadiabatic[2]—in scaling behaviour of DW dynamics arising from the competition. The contour maps are in excellent agreement with predictions of a generalized scaling theory, and all experimental data collapse onto a single curve. This result indicates that the adiabatic STT becomes dominant for large current densities, whereas the nonadiabatic STT—playing the same role as a magnetic field—subsists at low current densities required to make emerging magnetic nanodevices practical.**

The STT-driven DW motion opens great opportunities toward next-generation digital devices[4,5]. To achieve its full potential for practical nanodevices, it is necessary to construct STT devices working at low current densities, where the effects from disorders[7-14] in real devices become increasingly important. However the competition of the STTs against disorders is poorly understood. Even the central question of what is the main driving force against disorders yet remains controversial; it is reported experimentally to be the adiabatic STT[3] or the nonadiabatic STT[16,17]. However, in the former experiment[3] performed very close to the Curie temperature, critical magnetic fluctuations make quantitative analysis difficult. In the latter experiments[16,17], purely current-driven DW motion is not observed and it is difficult to distinguish Joule heating effects from other nonlinear effects that we present below. For an unambiguous experimental investigation, therefore, purely current-driven DW motion should be realized at temperatures far below the Curie temperature.

For this study, metallic Pt/Co/Pt films with the Curie temperature far above the room temperature are used. When their growth condition is optimized (Method Section), they exhibit clear circular domain expansion under a magnetic field as low as a few tenth of mT, which implies weak disorders in these films. Several nanowires are patterned on the films with different widths ranging from 190 to 470 nanometres (Fig. 1 with schematic electric connections). Here we show results from the 280-nm-wide nanowire and results from the others are shown in Supplementary Information. For the DW speed $V$ measurements, a DW is first created by the local Oersted field in the vicinity of the left vertical current line and then, pushed to a side by applying a magnetic field $H$ and/or a current density $J$ through the nanowire. The DW arrival time at the red-circled position (Fig. 1) is measured by the magneto-optical Kerr effect (MOKE) signal[12]. About four orders of magnitude in $V$ from $10^{-7}$ to $10^{-3}$ m/s are examined.

The field-driven DW motion is well understood and provides a useful starting point for the study of current-driven motion. The field-driven motion exhibits the creep scaling[10-14] $V(H)=V_0\exp(-\alpha H^{-\mu}/T)$ with the characteristic speed $V_0$, a scaling constant $\alpha$, and the temperature $T$. Note that in this Arrhenius-law-type formula, the energy barrier diverges as $H^{-\mu}$ since the competition between the field and disorder makes the motion collective. Figure 2 shows that our samples indeed follow the creep scaling with the exponent $\mu=1/4$ and $V_0=(2.6\pm0.4)\times10^4$ m/s.

To investigate roles of the STTs, we first examine effects of $J$ on the field-driven creep. To compensate for the Joule heating[18], $V$ (measured at the elevated temperature $T$) is converted into $V^*$ (speed at the constant ambient temperature $T_0$). Since the temperature rise $T-T_0$ is only a few degrees (Supplementary Information) over the experimental range of $J$, the conversion is achieved by $V^*=V_0\exp[(T/T_0)\ln(V/V_0)]$ within the scope of the Arrhenius law[17,19]. The circular symbols (Fig. 3a) show the typical results of $V^*$ with respect to $H$ for $\pm J$ ($J=7.7\times10^{10}$ A/m$^2$). The two curves clearly visualize the effect of $J$ on $V^*$. Interestingly, the two curves overlap onto each other (cross symbols), when shifted by proper amount $\pm\varDelta H_1$ in the horizontal direction, respectively. Figure 3b shows that the relation between $\varDelta H_1$ and $J$ is linear i.e., $\varDelta H_1=\varepsilon J$. We attribute this linear relation to the nonadiabatic STT[2], of which effect is known to be similar to the field. The coefficient $\varepsilon$ corresponds to the STT efficiency determined in depinning experiments[17,20,21]. For various Co/Pt-based systems, the reported STT efficiencies are 0.2 (Pt/0.6-nm Co/Pt[20]), 0.6 (Pt/[0.6-nm Co/Pt]$_3$[21]), 3.6 (Pt/[0.5-nm Co/Pt]$_2$[17]), and 8.0 (Pt/0.6-nm Co/AlO$_x$[20]) $\times10^{-14}$ Tm$^2$/A. The wide dispersion may be due to the spin current polarization $P$, which depends on the composition of Co and Pt[22]. The value of $|\varepsilon|$ in our Pt/Co/Pt nanowires is estimated to be $(1.6\pm0.1)\times10^{-14}$ Tm$^2$/A irrespective of the wire width. Since $|P|\leq1$, the nonadiabaticity $\beta$ is estimated[21] to be $\beta\gtrsim0.38$, which is comparable in magnitude to the Gilbert damping constant in Pt/Co/Pt films[23].

We now examine Fig. 3a more closely. Note that the cross symbols lie above the dotted line representing the purely field-driven speed. This implies that there exists an effect nonlinear in $J$, which boosts the speed regardless of the sign of $J$. One possible source of such an effect is the Joule heating. But we exclude this possibility since the Joule heating has been already taken care of in $V^*$. Moreover the attribution of the effect to the Joule heating leads to very unphysical conclusions (Supplementary Information).

To quantitatively examine such nonlinear effect, we construct a two-dimensional map of the speed $V^*(H,J)$ with respect to $H$ and $J$ (Fig. 4a). The colour corresponds to the magnitude of $V^*$. Each colour contrast thus visualizes the 'equi-speed' contour. For a quantitative analysis, the magnetic fields $H$ for several fixed $V^*$ are measured for each $J$ and plotted onto the map (symbols with error bars). It is interesting to see that, for each fixed $V^*$, $H$ is well characterized by quadratic functions of $J$, as shown by the contour lines of the best fit with the equation $H=H^*+\varepsilon J+cJ^2$, where $H^*$ is the value of the magnetic field at which each contour line crosses the magnetic field axis ($J=0$). Moreover we find that $c$ is proportional to $(H^*)^{1/2}$ (Fig. 4b). Incorporating all these observations, we construct the relation

$$H = H* + \varepsilon J + \eta\sqrt{H*}J^2 , \tag{1}$$

where $\eta$ is a proportionality constant.

Such quadratic contour lines agree with predictions of a generalized creep theory. Two core tasks of the theory is to determine the collective length $L_{col}$ over which the DW motion is correlated, and to evaluate the energy barrier $E(L_{col})$ that the DW segment of length $L_{col}$ has to overcome for the DW motion. Since $V=V_0\exp[-E(L_{col})/k_BT]$, the condition of constant $E(L_{col})$ amounts to the "equi-speed" condition.

According to Ref. 24, the energy barrier of general DW segment length $L$ is given by $E(L)=\varepsilon_{el}q^2L^{-1}-M_St_f(H-\beta P\chi J)qL+M_St_f\lambda P\chi J\psi L$, where $q$ and $\psi$ represent respectively the roughness and magnetization tilting angle of the segment. $L_{col}$ and $E(L_{col})$ are then determined from the condition $dE(L)/dL=0$ at $L=L_{col}$. Here, the first term of $E(L)$ represents the elastic energy, the second term is the combination of the magnetic Zeeman energy and the effective energy from the nonadiabatic STT, and the third term is the effective energy from the adiabatic STT. The physical meanings of the constants $\varepsilon_{el}$, $M_S$, $t_f$, $\lambda$, $\chi$ are given in the Supplementary Information. For metallic ferromagnets, it is well known[10-14] that the competition between the elastic energy and disorder makes $q$ proportional to $L^\zeta$ with $\zeta=2/3$. On the other hand, $\psi$ is reported[24] to be independent of $L$. However, we find that $\psi$ should be proportional to $J$ due to the competition between the magnetic anisotropy and the adiabatic STT (Supplementary Information).Then, the condition of the equal energy barrier leads to the equation

$$H^* \cong H - \varepsilon J - \eta J^2 \sqrt{H-\varepsilon J} + \tfrac{2}{5}\eta^2 J^4 + \cdots,\qquad(2)$$

which agrees, upon simple rearrangement, with the "equi-speed" condition in Eq. (1). This derivation reveals that $\varepsilon$ is proportional to the nonadiabaticity $\beta$ and the spin polarization $P$. It also reveals that the constant $\eta$ arises from the adiabatic STT. Thus, the nonlinear contribution in Eq. (1) is due to the adiabatic STT. The coefficient $\eta$ is estimated to be $(1.8\pm0.2)\times10^{-24}$ $T^{1/2}m^4/A^2$ for all the nanowires. Based on these experimental values, one find that the nonlinear contribution becomes larger than the linear contribution when $J>2\times10^{11}$ $A/m^2$.

An intriguing question is whether the relation in Eq. (2) holds even for the purely current-driven ($H=0$) motion. The left (right) panel in Fig. 5a shows the line-scanned MOKE images with a single (two) DW. Successive images—taken after each current pulse—indicate that the DW motion direction is independent of the magnetization direction of domains, but solely determined by the current direction. In our nanowires, such purely current-driven motion is accomplished at current densities less than $10^{11}$ A/m$^2$ (260 µA in total current through the device). Figure 5b shows $V^*$ versus $J$. The symbols with error bars represent the statistics from 50 measurements. The DW displacements are fairly well reproduced in measurements repeated up to several thousand times. Figure 5c plots both the purely current-driven DW motion (red symbols) and the purely field-driven DW motion (black symbols) as a function of $H^*$ defined in Eq. (2). Note that the two sets of data exactly overlap onto each other (Supplementary Fig. 6 for more comprehensive data). This verifies that Eq. (2) holds even for the purely current-driven DW motion. This also indicates that $J$ may be converted into an equivalent field $H^*$ through Eq. (2).

Lastly we discuss other effects that may affect the DW motion. The Oersted field from the current is irrelevant to the DW motion, since the two DWs (Fig. 5a) move in the same direction irrespective of the magnetic polarities of the neighbouring domains. The hydromagnetic drag and Hall charge effects[25] are estimated to be several orders smaller than our experimental data. Since interface effects become important in thin films, sizable Rashba spin-orbit coupling effect[26] is reported for strongly asymmetric layer structures Pt/Co/AlO$_x$. However this effect is negligible in our (almost) symmetric layer structure Pt/Co/Pt. Another interface effect is the renormalization of $P$. For Pt/Co interfaces, it is well known that the interface spin polarization is negative[22]. In ultrathin films like ours with the Co layer thickness of 0.3 nm, the interface spin polarization governs the device spin polarization $P$. For negative $P$, the DW should move, according to the STT theories[1,2], in the direction of current (rather than the electron motion direction), which is indeed the case in our sample (Fig.

5a). Except for this direction change, the sign reversal of $P$ does not affect the DW motion.

To conclude, our observation reveals unambiguously distinct roles of the adiabatic and nonadiabatic STTs in the competition against disorders. Unlike ideal cases without disorders, the adiabatic STT also plays a significant role for the DW motion. Our demonstration of the STT effects at low current densities will enhance opportunities toward low-power magnetic nanodevices.

**Methods**

The optimal properties of Si/100-nm $SiO_2$/5.0-nm Ta/2.5-nm Pt/0.3-nm Co/1.5-nm Pt ultrathin magnetic films are achieved by reducing the deposition rate as low as possible (~0.25 Å/sec) through adjustments to Ar sputtering pressure (~2 mTorr) and sputtering power (~10 W) to enhance the interface sharpness. In addition, the thicknesses of adhesion and protection layers are optimized to reduce disorders without ruining high perpendicular magnetic anisotropy. The base pressure is kept lower than $2 \times 10^{-8}$ Torr.

Nanowires with different widths, 190, 280, 360, and 470 nm are fabricated from the films by means of electron-beam lithography and ion milling. A negative tone electron-beam resist (maN-2403) is used for the lithography with fine resolution (~5 nm). To minimize the wire edge roughness, two ion milling steps are adopted with the different incident angles (15 and 75º) of ion beam. For the current injection and DW formation, two electrodes with coplanar waveguide geometry are stacked on each nanowire as shown in Fig. 1. To make an Ohmic contact, the surface of the nanowire is cleaned by $O_2$ plasma and weak ion milling before the electrode deposition.

A DW is created by the Oersted field[7] from a current pulse (33 mA, 1 μs) through the current line from the Function Generator 1 (FG1) to the ground, after saturating the magnetization at a magnetic field pulse (30 mT, 500 ms). Once a DW is formed in the vicinity to the current line, the DW is pushed to a side by applying magnetic field pulse and/or current pulse through the nanowire. The DW arrival time at a position (red circle), 15 μm away from the initial DW position is measured by use of a scanning MOKE microscope[12]. The current pulse profile is measured by an oscilloscope (OSC) connected in series between the nanowire and the ground. The current density through the nanowire is then estimated from the pulse amplitude with an assumption of uniform current distribution inside the nanowire, based on the fact that the total layer thickness is much smaller than the mean free path of the electrons. Every measurement is repeated by 50 times. The error bars in all the plots of the DW speed indicate the standard deviation.

**Acknowledgements**

This work was supported by the National Research Foundation of Korea grant funded by the Korea government (2007-0056952, 2009-0084542, 2008-0062257, 2009-0083723). JCL and KHS were supported by the KIST Institutional Program, by the KRCF DRC program, and by the IT R&D program of MKE/KEIT (2009-F-004-01).

Correspondence and requests for materials should be addressed to S.-B.C. and H.-W.L.

**Competing financial interests**

The authors declare that they have no competing financial interests.

Figure Legends

**Figure 1 | Typical device structure with electric connections.** Secondary electron microscope (SEM) image for a nanowire with schematic drawing of measurement setup. The definition of the positive polarities of $J$ and $H$ is shown in the inset.

**Figure 2 | Creep plot of $V$ with respect to $H^{-1/4}$.** The symbols show averaged values of 50 repeated measurements for each $H$ and the error bars (ordinate) are the standard deviation. The error bars (abscissa) are the maximum inaccuracy of the $H$ measurement. The line shows the best linear fit.

**Figure 3 | Field-driven DW motion under bias current density.** (**a**) Relation between $V^*$ and $H$, for $J=\pm7.7\times10^{10}$ A/m$^2$ (olive/purple circular symbols), respectively. The symbols show the averaged values of 50 repeated measurements for each $H$ and $J$, and the error bars (ordinate) are the standard deviation. The error bars (abscissa) are the maximum inaccuracy of the $H$ measurement. The cross symbols show the data shifted in the horizontal $H$ axis by $\pm\Delta H_1$ (=$\pm1.2$ mT), accordingly. The black dotted line exhibits the purely field-driven DW speed, identical to the best fit in Fig. 2. (**b**) $\Delta H_1$ with respect to $J$. The error bars are the maximum inaccuracy of the $\Delta H_1$ and $J$ measurements. The dotted line shows the best linear fit.

**Figure 4 | Contour map of DW speed $V^*$ with respect to $H$ and $J$.** (**a**) The colour corresponds to the magnitude of $V^*$ as given by the colour bar. The symbols indicate the magnetic fields $H$ for several fixed $V^*$ for each $J$. The error

bars are determined by the standard deviation in $V^*$ measurements divided by the slope d$V^*$/d$H$. The solid lines are the best fit with Eq. (1). (**b**) Scaling plot between $c$ and $H^*$. The error bars are the standard deviations for $c$ and $H^*$ for several repeated measurements. The dotted line shows the best linear fit.

**Figure 5 | Purely current-driven DW motion.** (**a**) Domain images for a DW (left) and two DWs (right) from the line-scanned MOKE signal, successively taken after each current pulse ($\pm 1.7 \times 10^{11}$ A/m$^2$, 1.5 ms). The arrows indicate the direction of the current flow. (**b**) $V^*$ vs $J$. The symbols show averaged values of 50 repeated measurements for each $J$ and the error bars (ordinate) are the standard deviation. The error bars (abscissa) are the maximum inaccuracy of the $J$ measurement. (**c**) Creep scaling plot of $V^*$ with respect to $H^*$ (red). The error bars are identical to those in (b). The black solid line and the black symbols with error bars are identical to those in Fig. 2.

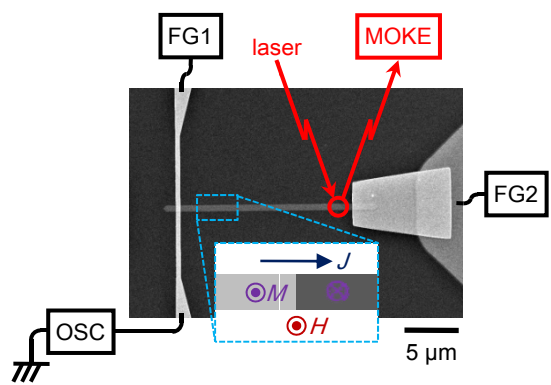

Figure 1

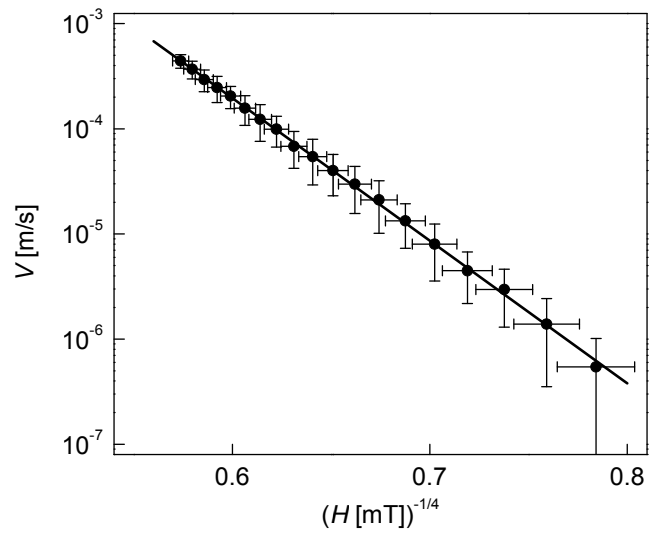

Figure 2

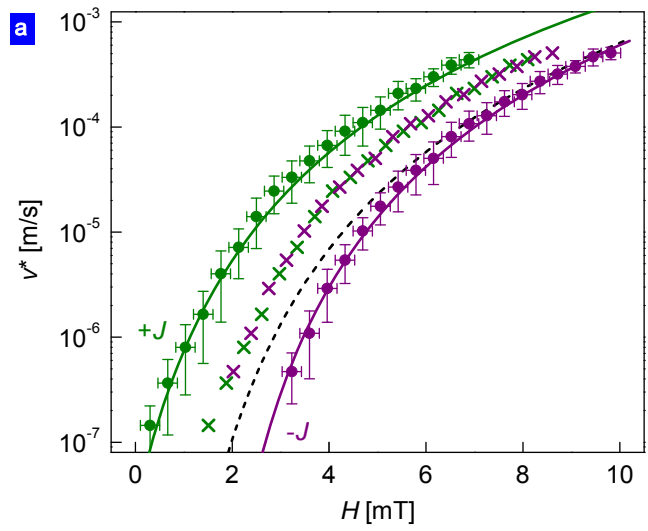

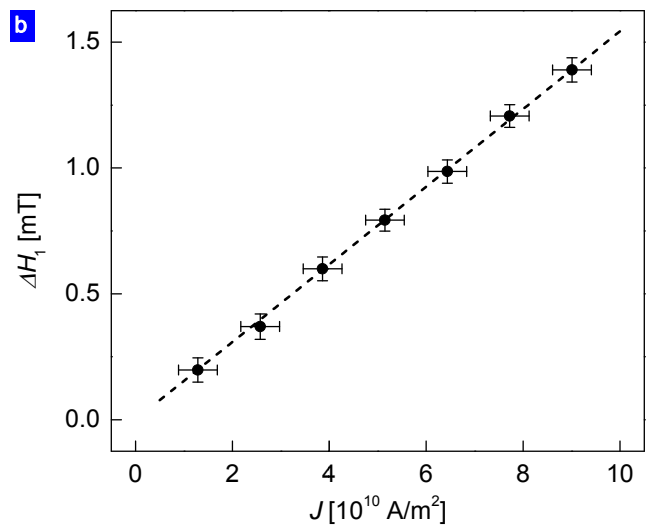

Figure 3

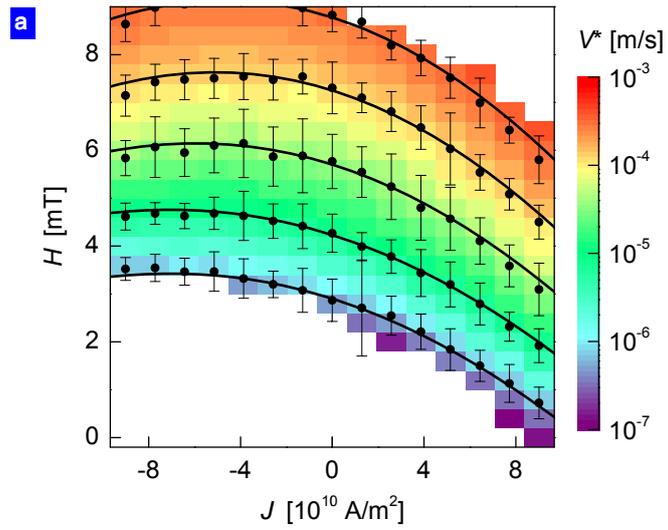

**a**

$V^*$ [m/s]

$H$ [mT]

$J$ [$10^{10}$ A/m$^2$]

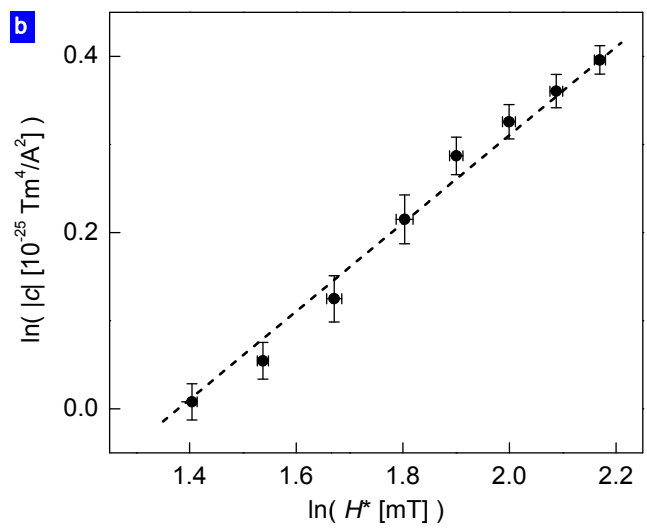

**b**

$\ln(\,|c|\,[10^{-25}\ \mathrm{Tm^4/A^2}]\,)$

$\ln(\,H^*\ [\mathrm{mT}]\,)$

Figure 4

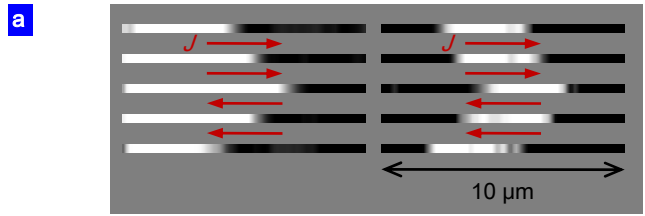

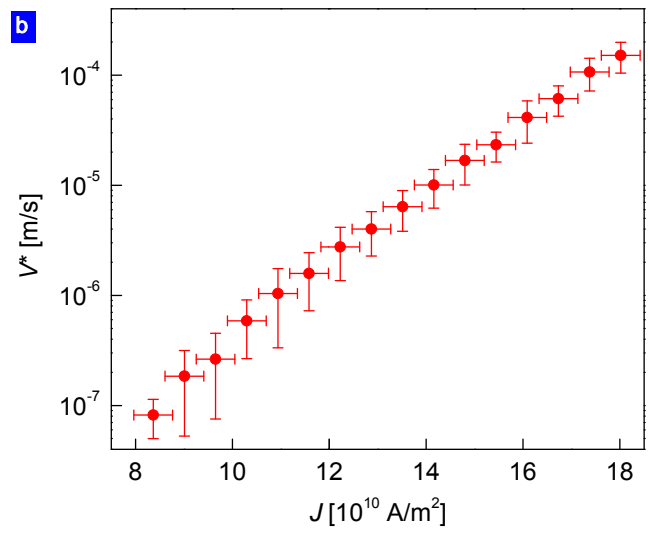

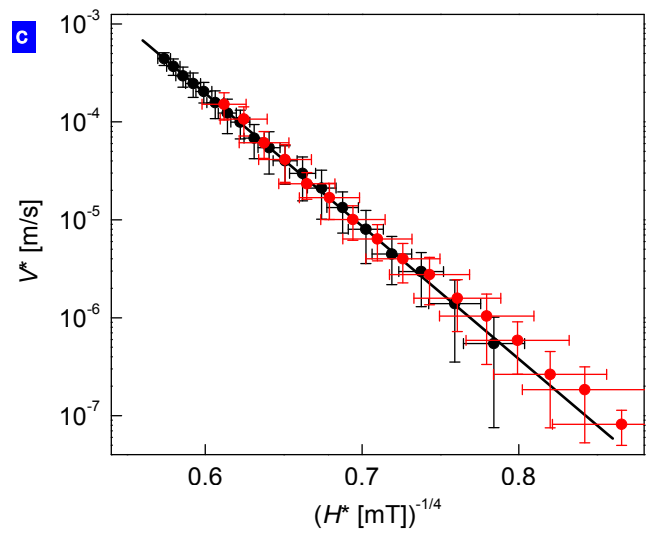

Figure 5

# Supplementary Methods

## I. Sample Preparation

Si/100-nm SiO$_2$/5.0-nm Ta/2.5-nm Pt/0.3-nm Co/1.5-nm Pt ultrathin magnetic films are chosen because these films exhibit fairly small propagation field and clear circular domain wall (DW) expansion. Such optimal properties are achieved by reducing the deposition rate as low as possible (~0.25 Å/sec) through adjustments to Ar sputtering pressure (~2 mTorr) and sputtering power (~10 W) to enhance the interface sharpness. In addition, the thicknesses of adhesion and protection layers are optimized to reduce disorders without ruining high perpendicular magnetic anisotropy (PMA). The base pressure is kept lower than $2 \times 10^{-8}$ Torr.

The films exhibit strong PMA, as evidenced by the hysteresis measurement using an alternating gradient magnetometer (AGM). The saturation magnetization is determined to be 1.6±0.2 T from the measurement. The temperature-dependent extraordinary Hall effect (EHE) measurement reveals that the Curie temperature is larger than the measurement range up to 350 K, by showing finite coercive field and remnant magnetization in hysteresis loops. From the angle-dependent EHE measurement[27], the anisotropy field is estimated to be about 1.3±0.1 T, which corresponds to the PMA constant $(8.3±0.2) \times 10^{5}$ J/m$^3$.

Nanowires with different widths, 190, 280, 360, and 470 nm are fabricated from the films by means of electron-beam lithography and ion milling. A negative tone electron-beam resist (maN-2403) is used for the lithography with fine resolution (~5 nm). To minimize the wire edge roughness, two ion milling steps are adopted with the different incident angles (15 and 75º) of ion beam. For the current injection and DW formation, two electrodes with coplanar waveguide geometry are stacked on each nanowire as shown in Fig. 1. To make an Ohmic contact, the surface of the nanowires is cleaned by $O_2$ plasma and weak ion milling before the electrode deposition.

## II. Measurement Procedure

To generate a DW in a nanowire, we first saturate the magnetization (pointing down to the sample plane) with a sufficiently high magnetic field pulse (30 mT, 500 ms) and then, inject a current pulse (33 mA, 1 μs) through the current line from the Function Generator 1 (FG1) to the ground as shown in Fig. 1. The current pulse generates an Oersted field (pointing up from the sample plane) at the right side of the current line. The maximum strength of the Oersted field is estimated to be about 30 mT in the vicinity of the current line, which is large enough to reverse the local magnetization (pointing up from the sample plane). A DW is then placed between the reversed (up) and unreversed (down) domains[7]. We confirm that the DW position is reproducibly placed less than 1 μm away from the current line. Once a DW is formed, the DW is pushed to a side by applying current pulse and/or magnetic field pulse. The magnetic field $H$ is applied to the direction of the magnetization in the reversed (up) domain and the current $J$ is applied from left to right by the Function Generator 2 (FG2), with the definition of the positive polarities of $H$ and $J$ shown in the inset. The DW arrival time at a position (red circle), 15 μm away from the initial DW position is measured by use of a scanning magneto-optical Kerr effect (MOKE) microscope[12]. An abrupt MOKE signal drop indicates the passage of a DW. To precisely measure the DW arrival time, we simultaneously measure both the MOKE signal and the trigger signal for the magnetic field and/or current. The current pulse amplitude and duration are measured by use of an oscilloscope (OSC) connected in series between the nanowire and the ground. The voltage $V_O$ measured by the oscilloscope is converted to the current $I$ by the relation $I=V_O/R_O$ with the load resistance $R_O$ (50 Ω) of the oscilloscope input. For this measurement, the FG1 is disconnected. The current density $J$ through the nanowire is then estimated by the relation $J=I/wt_f$, where $w$ is the wire width and $t_f$ is the total film thickness (9.3 nm). This estimation is done with an assumption of a uniform current distribution inside the nanowire, based on the fact that the total layer thickness is much smaller than the intrinsic mean free path of the electrons. Every measurement is repeated 50 times. The error bars in all the plots of the DW speed indicate the standard deviation.

## Supplementary Discussions

### III. DW Propagation in Films

The magnetic domain expands circularly as shown by the MOKE image in Supplementary Fig. 1a. The image is obtained by accumulating the domain images taken at successive times with a constant time step (1 s) under an external magnetic field (0.8 mT). Each gray contrast thus exhibits the domain patterns, where darker area corresponds to the domain pattern at earlier time. The formation of the clear circular domain patterns with less jaggedness manifests that the magnetization reversal is dominated by the DW motion through weak microstructural disorders. Such domain propagation is observed even under a weak magnetic field much less than 1 mT. From the clear circular DW expansion, one can easily measure the DW speed $V$ as a function of the external magnetic field $H$. Supplementary Fig. 1b shows the creep scaling plot of $V(H)$, where the abscissa scales as $H^{-1/4}$. The DW propagation obeys the creep scaling as shown by the linear dependence in the scaled axes.

## IV. Temperature Measurement

The temperature rise due to the Joule heating is estimated via the temperature-dependent electric resistivity of nanowires[18]. In this method, the resistivity $\rho$ of a nanowire is measured with respect to the current density $J$ as shown by Supplementary Fig. 2a. Due to the Joule heating, the resistivity exhibits a quadratic dependence on the current density as shown by the best fit (solid line) with the equation $\rho/\rho_0 = 1 + \sigma_J J^2$, where $\rho_0$ is the resistivity at the ambient temperature (297 K) and $\sigma_J$ is a proportionality constant. The measured resistivity is converted to the temperature rise $\Delta T$ by use of the relation $\rho/\rho_0 = 1 + \sigma_T \Delta T$ with a proportionality constant $\sigma_T$, which is preliminarily measured with respect to the temperature $T$ in a cryostat as shown by Supplementary Fig. 2b. Combining these two measurements, we estimate the temperature rise as $\Delta T = (\sigma_J/\sigma_T) J^2$. In experiments, $\sigma_T$ and $\sigma_J$ are estimated to be $3.7 \times 10^{-4}$ K$^{-1}$ and $2.5 \times 10^{-25}$ (A/m$^2$)$^{-2}$, respectively. For the current and field-driven measurement (Figs. 3 and 4) up to $J = 9.0 \times 10^{10}$ A/m$^2$, $\Delta T$ stays less than 6 K and for the current-only measurement (Fig. 5b) up to $J = 1.8 \times 10^{11}$ A/m$^2$, $\Delta T$ reaches about 23 K.

## V. Existence of Nonlinear Contribution in Effective Field

As discussed in the manuscript, Figs. 3a (deviation between the cross symbols and the dotted line) and 4a (curved equi-speed lines) indicate that the current induces not only the linear contribution $\Delta H_1(J)$ but also a nonlinear contribution to the effective magnetic field. Here we show that those features in Figs. 3a and 4a cannot be ascribed to the Joule heating effect. To demonstrate this, we use an inductive reasoning method (reductio ad absurdum); we assume that the effective field due to the current density does not contain any nonlinear contribution in $J$ and that the features in Figs. 3a and 4a are due to the Joule heating. Below we demonstrate that these assumptions lead to unreasonable results. We first remark that $\varepsilon$ determined from Fig. 3a is not affected by the Joule heating effect since the same amount of the Joule heating is expected for $\pm J$. Then the total effective field is given by $H+\Delta H_1(J)= H+\varepsilon J$ since the nonlinear contribution of $J$ to the effective field is assumed to be absent. Supplementary Fig. 3 shows the $V$ vs. $H+\varepsilon J$ relation for $J=\pm 9.0\times10^{10}$ A/m$^2$ (olive, purple), in comparison with the corresponding relation for $J=0$ A/m$^2$ (black), for the 280-nm-wide nanowire. It is evident that the slope of the olive solid line (for $|J|=9.0\times10^{10}$ A/m$^2$) is different from the slope of the black solid line (for $J=0$ A/m$^2$) with the deviation clearly above the error bars. One possible origin of the deviation is the temperature rise due to the Joule heating, since the slope is inversely proportional to the temperature $T$. This explanation however requires the temperature rise of about 60 K for $|J|=9.0\times10^{10}$ A/m$^2$ to account for the slope difference, while the temperature rise estimated from the electrical resistance measurement is less than one tenth of the required value. This shows that the features in Figs. 3a and 4a cannot be attributed to the Joule heating. Supplementary Fig. 3 reveals another absurd implication of the assumptions. Note that the intercepts to the ordinate, which correspond to $V_0$, are also quite different (about factor 10 difference) depending on $|J|$. It clearly violates the criterion of the creep scaling since, in creep scaling, $V_0$ is given by a characteristic constant[10-14] irrespective of $H$ and $T$. Therefore, one can conclude that the above assumptions are not valid and there exists a nonlinear contribution of the current to the effective field.

## VII. Generalized Creep Formula

The free energy $E$ of a DW segment with the length $L$ is given by a function of the roughening amplitude $q$ and the tilting angle of the magnetization $\psi$, as[24]

$$E = \varepsilon_{el}\frac{q^2}{L} - M_S t_f\left(H - \beta P\chi J\right)qL + M_S t_f \lambda P\chi J\psi L \,, \tag{S1}$$

with the elastic energy density $\varepsilon_{el}$, the saturation magnetization $M_S$, the film thickness $t_f$, the wall width $\lambda$, and the spin current polarization $P$. Here, $\chi$ is the conversion parameter from $J$ to the magnetic field dimension[20,21]. According to the discussion after Eq. (16) in Ref. 24, $\psi$ is proportional to $J$ and thus, we denote $\psi = \psi_0 J$ with a proportionality constant $\psi_0$. In real films with disorders, $q$ follows a scaling law $q = q_0(L/L_C)^\zeta$ with the wandering exponent $\zeta$, where $q_0$ and $L_C$ are the scaling constants. The free energy $E$ is then written as

$$E(L) = \varepsilon_{el}\frac{q_0^2}{L_C^{2\zeta}}L^{2\zeta-1} - M_S t_f\left(H - \beta P\chi J\right)\frac{q_0}{L_C^\zeta}L^{\zeta+1} + M_S t_f \lambda P\chi J^2\psi_0 L \,. \tag{S2}$$

For metallic ferromagnets[10-14] with $\zeta=2/3$, the free energy can be rewritten as

$$E(L) = AL^{1/3} - BL^{5/3} + CL, \tag{S3}$$

where $A=\varepsilon_{el}q_0^2L_C^{-4/3}$, $B=M_S t_f(H-\beta P\chi J)q_0 L_C^{-2/3}$, and $C=M_S t_f \lambda P\chi \psi_0 J^2$. The maximum free energy is then determined by

$$\left.\frac{\partial E}{\partial L}\right|_{L_{col}} = \frac{1}{3}AL^{-2/3} - \frac{5}{3}BL^{2/3} + C = 0. \tag{S4}$$

From Eq. (S4), the collective length[12] $L_{col}$ is given by

$$L_{col} = \left(\frac{-3C + \sqrt{9C^2 + 20AB}}{2A}\right)^{-3/2}, \tag{S5}$$

and then, the energy barrier $E_B$ i.e. the maximum free energy is written as

$$E_B = E(L_{col}) = \frac{2}{5}(2A)^{3/2}\frac{\left(-2C + \sqrt{9C^2 + 20AB}\right)}{\left(-3C + \sqrt{9C^2 + 20AB}\right)^{3/2}}. \tag{S6}$$

Replacing $B$ by $D(H-\varepsilon J)$ where $D=M_S t_f q_0 L_C^{-2/3}$ and $\varepsilon=\beta P\chi$, Eq. (S6) becomes

$$E_B = \frac{2}{5}\frac{(2A)^{3/2}}{(20AD)^{1/4}}\frac{\left(-\dfrac{2C}{\sqrt{20AD}} + \sqrt{\dfrac{9C^2}{20AD} + (H-\varepsilon J)}\right)}{\left(-\dfrac{3C}{\sqrt{20AD}} + \sqrt{\dfrac{9C^2}{20AD} + (H-\varepsilon J)}\right)^{3/2}}, \tag{S7}$$

and then, it can be written as $E_B=(2/5)(2A)^{3/2}(20AD)^{-1/4}\{F(H,J)\}^{-1/4}$ with

$$F(H,J) = \frac{\left(-3\eta J^2/10 + \sqrt{\left(3\eta J^2/10\right)^2 + (H-\varepsilon J)}\right)^6}{\left(-\eta J^2/5 + \sqrt{\left(3\eta J^2/10\right)^2 + (H-\varepsilon J)}\right)^4}, \tag{S8}$$

by replacing $3C/(20AD)^{1/2}=3\eta J^2/10$ where $\eta=\psi_0\lambda L_C P\chi(5M_S t_f/\varepsilon_{el}q_0^3)^{1/2}$. Since the DW speed is determined solely by $E_B$, the 'equi-speed' contour amounts to the condition of a constant $E_B$ or a constant $F(H,J)$. Since $F(H^*,0)=H^*$, the 'equi-speed' contour that pass through the point ($H=H^*,J=0$) is determined by the equation $H^*=F(H,J)$. The Taylor expansion of (S8) with respect to $\eta J^2$ is written as

$$H^* = F\left(H,J\right) = H - \varepsilon J - \eta J^2 \sqrt{H - \varepsilon J} + \frac{2}{5}\left(\eta J^2\right)^2 + O[J]^6, \qquad (S9)$$

in accordance with Eq. (2). The inverse function of Eq. (S9) can be obtained by solving the equation $(X+2/3)^2=Y(X+1)^3$, where $X=\{1+(H-\varepsilon J)/(3\eta J^2/10)^2\}^{1/2}$ and $Y=-3\eta J^2/10(H^*)^{1/2}$, for the case of $\eta<0$. The solution is given by

$$X = \frac{1-3Y}{3Y} + \frac{2\sqrt{1-2Y}}{3Y}\cos\left[\frac{1}{3}\tan^{-1}\left(\frac{\sqrt{4Y^3 - 9Y^4}}{2 - 6Y + 3Y^2}\right)\right]. \qquad (S10)$$

Since the value of the cosine in the equation is ~1 when $Y<0.3$ as in our experiments, the solution becomes $X\cong\{1+2(1-2Y)^{1/2}-3Y\}/3Y$. Replacing $X$ and $Y$ by the original definition, one obtains

$$H \cong \varepsilon J + \frac{H^*}{9}\left\{5 + \frac{21}{5}\frac{\eta J^2}{\sqrt{H^*}} + 4\left(1 + \frac{9}{10}\frac{\eta J^2}{\sqrt{H^*}}\right)\sqrt{1 + \frac{3}{5}\frac{\eta J^2}{\sqrt{H^*}}}\right\}. \qquad (S11)$$

The Taylor expansion of Eq. (S11) is written as

$$H \cong H^* + \varepsilon J + \eta\sqrt{H^*}J^2 + \frac{1}{10}\eta^2 J^4 + O[J]^6, \qquad (S12)$$

which accords with Eq. (1).

## VI. Contour Maps and Universal Curves of Other Nanowires

We plot the contour maps of other nanowires with different width; (a) 190, (b) 360, and (c) 470 nm in Supplementary Fig. 4, respectively. All the nanowires also exhibit clearly the quadratic relation between $H$ and $J$ on the 'equi-speed' contour. The linear contribution to the effective field $\Delta H_1$ is summarized in Supplementary Fig. 5a for all the nanowires, where the linear coefficient $|\varepsilon|$ is estimated to be almost unchanged with respect to the wire width: 1.67 (190 nm), 1.54 (280 nm), 1.43 (360 nm), and 1.71 (470 nm) $\times 10^{-14}$ $Tm^2/A$, respectively. The quadratic coefficients $c$ for all the nanowires are confirmed to be proportional to the square-root of $H^*$, as shown in Supplementary Fig. 5b. The adiabatic coefficient $\eta$ is determined to be 1.3 (190 nm), 1.7 (280 nm), 1.8 (360 nm), and 2.1 (470 nm) $\times 10^{-24}$ $T^{1/2}m^4/A^2$, respectively. Note that the adiabatic coefficient $\eta$ is determined from the contour maps by use of Eq. (S11) rather than Eq. (1). Based on these experimentally-determined values of $\varepsilon$ and $\eta$, the universality in the DW motions driven by either $J$ or $H$ (or even both) is confirmed as shown in Supplemental Fig. 6a-d.

## VII. Notes for Scaling Exponents

We examine the possibility that the current-driven DW motion is described by a simpler formula $V=V_0\exp(-bJ^{-\mu}/T)$ with an exponent possibly different from the field-driven DW motion exponent 1/4. Supplementary Fig. 7 shows the relation between $V^*$ and $J^{-\mu}$ with several different exponents $\mu$, (a) the field-driven exponent[10-14] 1/4, (b) the exponent 1/3 experimentally observed in ferromagnetic semiconductor[3] (Ga,Mn)As, (c) the exponent 1/2 theoretically proposed for the DW motion with the adiabatic STT[3], and

(d) the exponent –1/2 from the best fit. It is peculiar that the best fit is achieved for a negative exponent. However the negative exponent violates the length hierarchy of the scaling theory of the creep as discussed in the Supplementary Section IV in Ref. 12. When the value of $\mu$ is confined to the physically allowed range, $\mu > 0$, it is evident that the experimental data exhibit systematic deviations (a-c) from the simpler formula regardless of $\mu$ values. This failure of the simpler formula may be ascribed to the coexistence of the two driving forces (adiabatic and nonadiabatic STTs) with comparable strengths. For complete understanding in the scaling relation, an analysis with consideration of the two-force competition has to be followed.

## Supplementary Notes

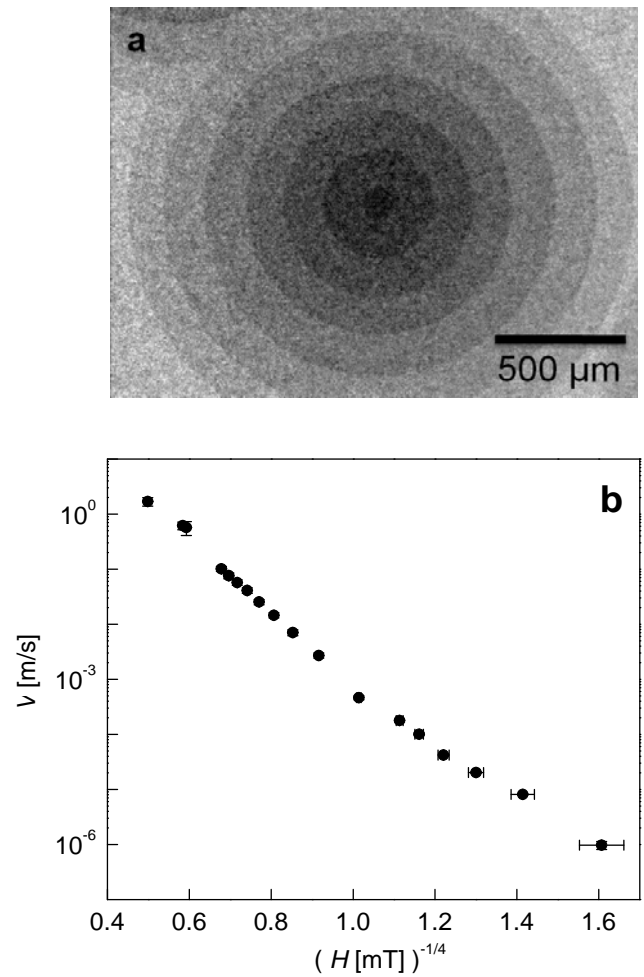

**Supplementary Figure 1** (a) Typical domain images measured by a polar MOKE microscope. The gray contrasts correspond to the successive domain images in time under an applied magnetic field (0.8 mT). The darker area corresponds to the domain pattern at earlier time. (b) Creep plot of the DW speed $V$ with respect to the applied field $H^{-1/4}$.

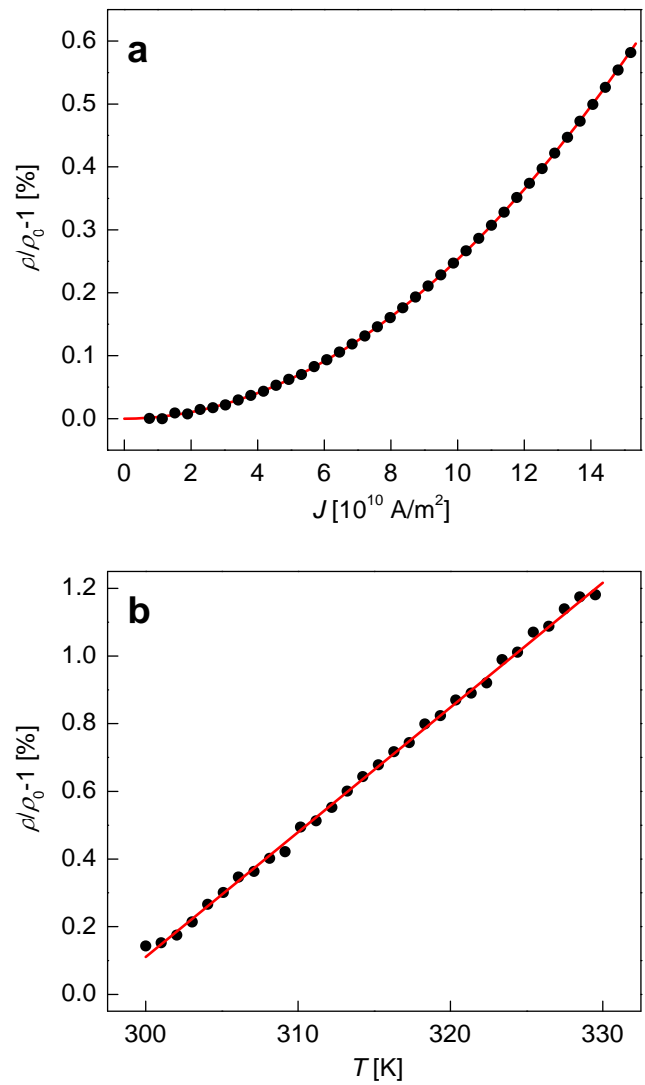

**Supplementary Figure 2** (a) The variation in resistivity $\rho/\rho_0$ with respect to the current density $J$ for the 280-nm-wide nanowire. The solid line is the best fit with the equation $\rho/\rho_0 = 1 + \sigma_J J^2$. (b) The variation in resistivity $\rho/\rho_0$ with respect to the temperature $T$ measured in a cryostat. The solid line is the best fit with the equation $\rho/\rho_0 = 1 + \sigma_T \Delta T$.

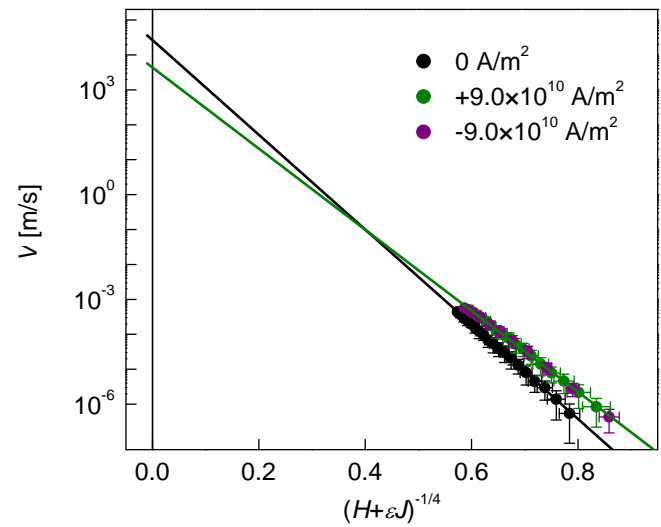

**Supplementary Figure 3** DW speed *V* with respect to the scaled axis (*H*+ε*J*)^−1/4 for the 280-nm-wide nanowire, under the assumption of the linear proportionality between the effective field and the current. The colour corresponds to different current bias, *J*=0 (black), +9.0 (olive), and −9.0 (purple) times 10^10 A/m^2, respectively. The solid lines are the best linear fits.

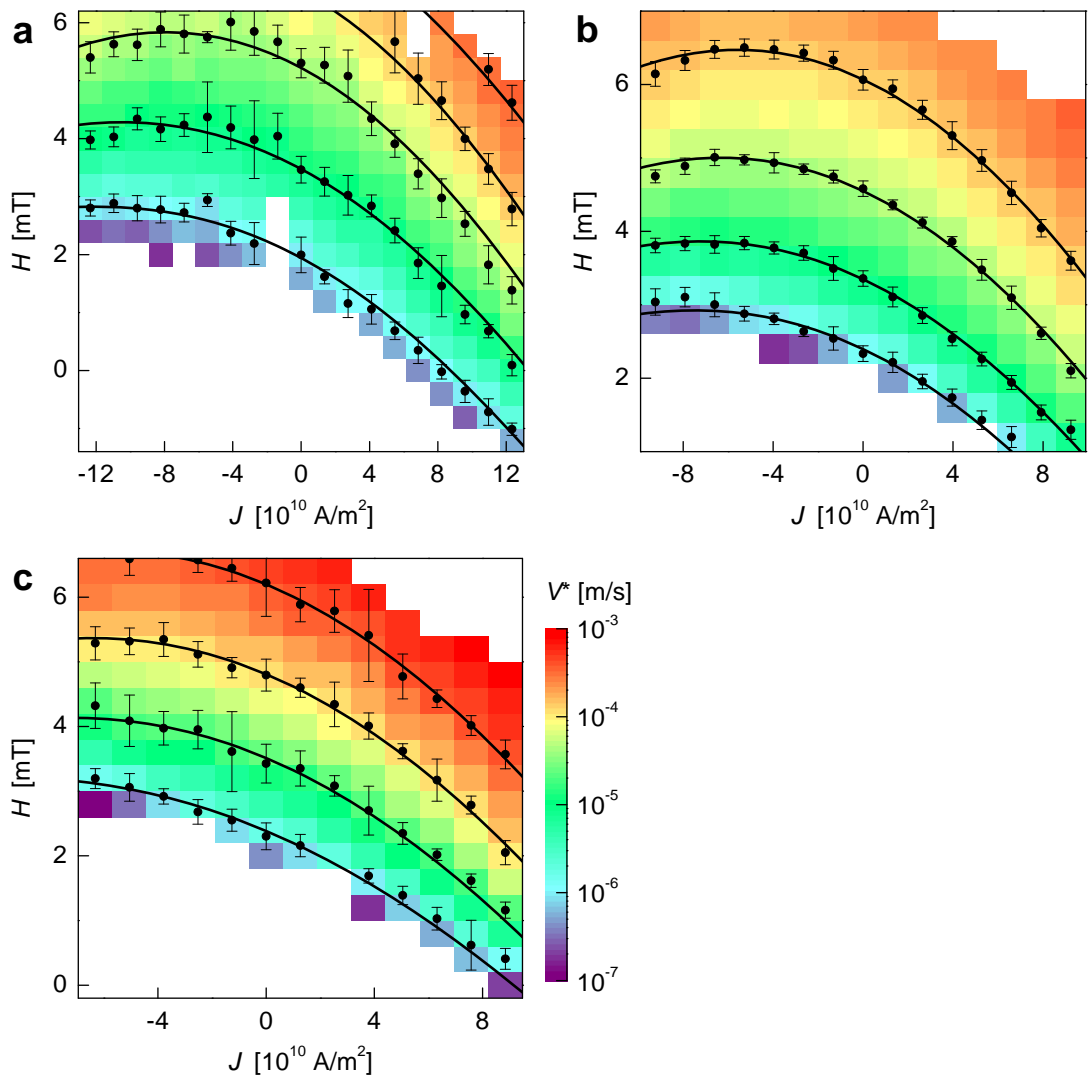

**Supplementary Figure 4** Contour plots for the nanowires with different widths, (a) 190, (b) 360, and (c) 470 nm, respectively. The symbols indicate the magnetic fields *H* for several fixed *V** for each *J*. The error bars are determined by the standard deviation in *V** measurements divided by the slope d*V**/d*H*. The solid lines are the best fit with Eq. (1).

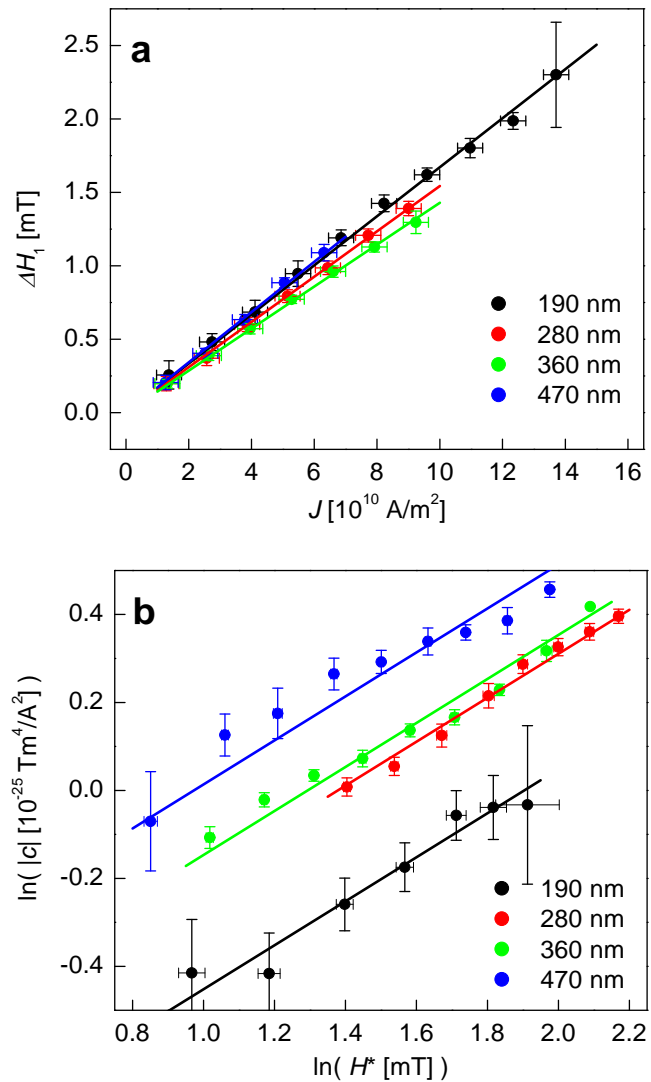

**Supplementary Figure 5** The parameters measured for the nanowires with different widths, (black) 190, (red) 280, (green) 360, and (blue) 470 nm, respectively. (a) The linear contribution to the effective field $\Delta H_1$ with respect to $J$. The lines are the best linear fits. (b) The log-log scaling plot of the quadratic coefficients $c$ with respect to $H^*$. The lines are the best linear fits with the slope 1/2.

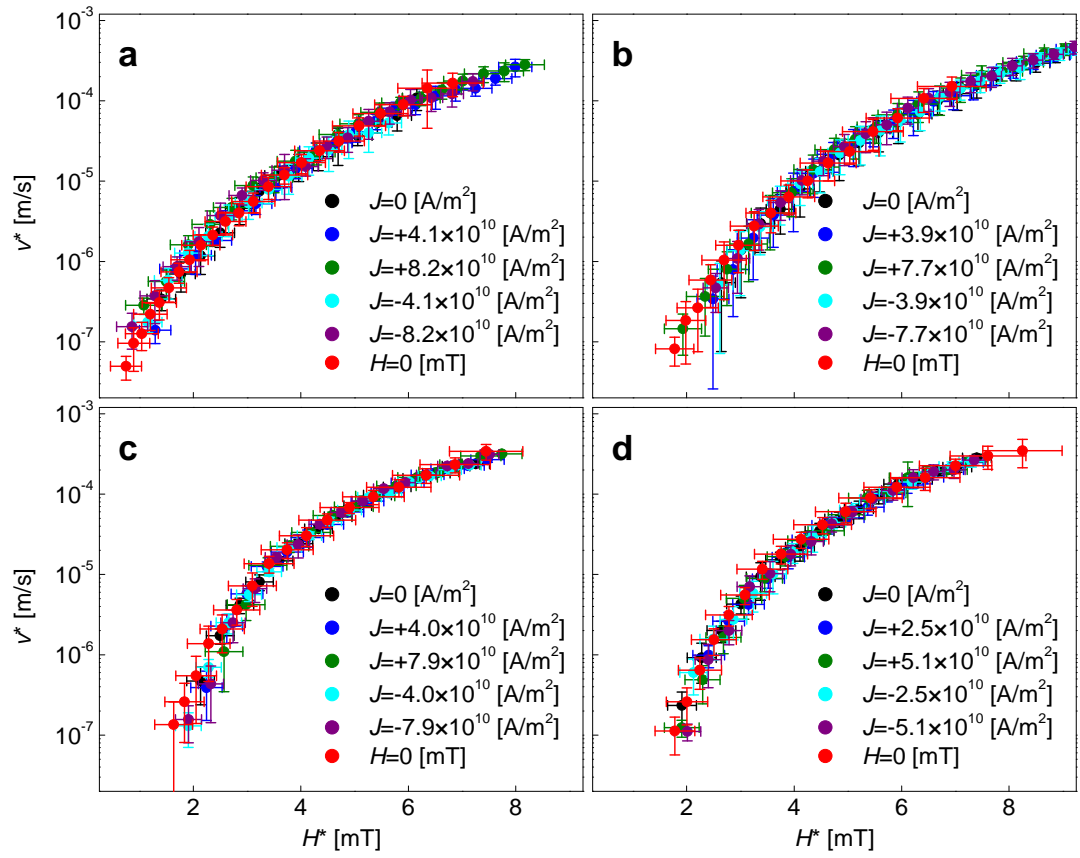

**Supplementary Figure 6** Plots of $V^*$ with respect to $H^*$ for (a) 190, (b) 280, (c) 360, and (d) 470 nm, respectively. Each plot shows data for all the DW motions driven purely by $J$ (red) or purely by $H$ (black), or jointly by both $J$ and $H$ (blue, olive, cyan, purple with different $J$ denoted in each plot).

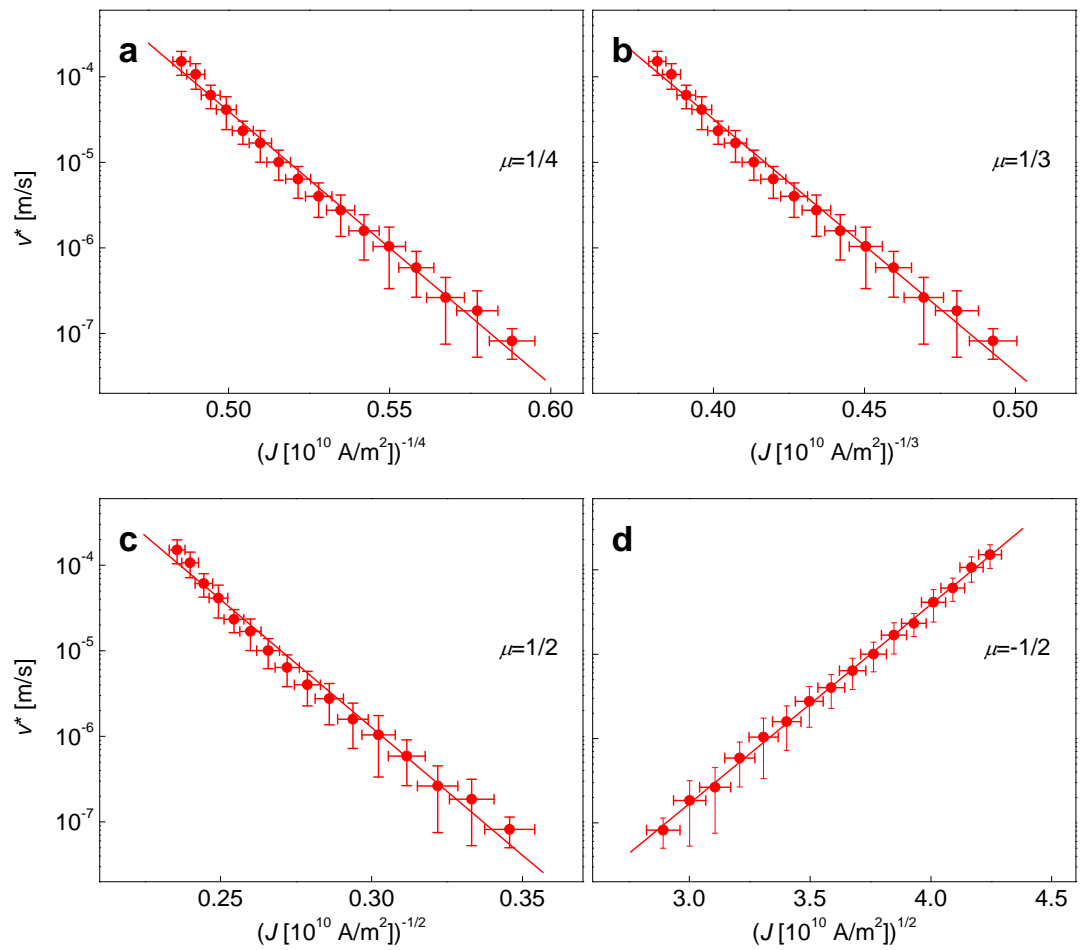

**Supplementary Figure 7** Scaling plots for several different exponents; (a) 1/4, (b) 1/3, (c) 1/2 and (d) −1/2, respectively. All the values with the error bars are identical to those shown in Figs. 5b.